\documentclass[ a4paper,floats,aps,nofootinbib,superscriptaddress,12pt,
tightenlines, preprintnumbers]{revtex4}

\usepackage{amsmath}
\usepackage{graphicx}
\usepackage{amsfonts}
\usepackage{amssymb}%
\usepackage{epsf}
\usepackage{epsfig}
\usepackage{hyperref}

\def\noi{\noindent}
\def\non{\nonumber}

\newcommand{\chargomp}{{\tilde{\chi}}_1^\mp}
\newcommand{\chargtpm}{{\tilde{\chi}}_2^\pm}

\newcommand{\tb}{t_\beta}

\newcommand{\cosb}{c_\beta}

\newcommand{\sinb}{s_\beta}
\newcommand{\sbtwo}{s_{2\beta}}

\newcommand{\Lag}{\mathcal{L}}

\newcommand{\ra}{\rightarrow}
\def\l{\lambda}

\def\b{\beta}
\def\g{\gamma}

\def\k{\kappa}

\begin{document}

\title{
Boosting Higgs decays into gamma and a Z in the NMSSM
}

\preprint{LAPTH-01/14}
\preprint{LPSC 14-017}

\author{Genevi\`{e}ve~B\'{e}langer}
\email[]{belanger@lapth.cnrs.fr}
\author{Vincent~Bizouard}
\email[]{bizouard@lapth.cnrs.fr}
\affiliation{LAPTh, Universit\'{e} de Savoie, CNRS, 9 Chemin de Bellevue, B.P.
110, F-74941 Annecy- le-Vieux, France}
\author{Guillaume~Chalons}
\email[]{chalons@lpsc.in2p3.fr}
\affiliation{LPSC, Universit\'{e} Grenoble-Alpes, CNRS/IN2P3,
Grenoble INP, 53 rue des Martyrs, F-38026 Grenoble, France}

\begin{abstract}
 In this work we present the computation of the Higgs decay into a
photon and a $Z^0$ boson at the one-loop level in the framework of the
Next-to-Minimal Supersymmetric Standard Model (NMSSM). The numerical evaluation
of this decay width was performed within the framework of the \texttt{SloopS}
code, originally developed for the Minimal Supersymmetric Standard Model (MSSM)
but which was recently extended to deal with the NMSSM. Thanks to the high level
of automation of \texttt{SloopS} all contributions from the various sector of
the NMSSM are consistently taken into account, in particular the non-diagonal
chargino and sfermion contributions. We then explored the NMSSM parameter
space, using \texttt{HiggsBounds} and \texttt{HiggsSignals}, to investigate to
what extent these signal can be enhanced.
\end{abstract}

\date{\today}
 \maketitle

\section{Introduction}
The discovery of a 125 GeV Higgs boson at the LHC in July 2012
\cite{Aad:2012tfa,Chatrchyan:2012ufa}, is a milestone in the road leading to the
elucidation the ElectroWeak Symmetry Breaking (EWSB) riddle. Since then, its
couplings to electroweak gauge bosons, third generation fermions and the
loop-induced couplings to photon and gluons have been measured with an already
impressive accuracy by the ATLAS and CMS collaborations
\cite{ATLAS:2013sla,CMS:yva} during the 7 and 8 TeV runs. This great achievement
was made possible since for a 125 GeV Higgs bosons many different production and
decay channels are detectable at the LHC. A spin and parity analysis of the
Higgs boson in the decays $H\ra \g\g$, $H \ra ZZ^*$ and $H\ra W W^*$ favor the
$J^P = 0^+$ hypothesis \cite{Aad:2013xqa,Chatrchyan:2012jja}. 
\par\noi
The couplings of the Higgs to photons $H\g\g$ and gluons $Hgg$ are
induced at the quantum level, even in the Standard Model (SM), and thus are
interesting probes of New Physics (NP) since both the SM and NP contributions
enter at the same level. On the one hand, the updated CMS
analysis \cite{CMS:ril} gives a signal strength for the
diphoton which is now compatible with the SM, compared to the previous one. On
the other hand, the updated ATLAS analysis \cite{ATLAS:2013oma} still observes a
slight excess of events in this channel but recent measurements of the $H\ra
\g\g$ differential cross sections do not show significant disagreements with
expectations from a SM Higgs \cite{ATLAS-CONF-2013-072}.
\par\noindent
The search for another important loop-induced Higgs decay channel, $H\ra \g
Z^0$, is also performed by the ATLAS and CMS experiments
\cite{Chatrchyan:2013vaa,ATLAS-CONF-2013-009}. Within the SM, the partial width
for this channel, $\Gamma_{\g Z}$, is about two thirds of that for the diphoton
decay and its measurement can also provide insights about the properties of the
boson, such as its mass, spin and parity \cite{Gainer:2011aa}, thanks to a
clean final state topology. No excess above SM predictions has been found in the
120-160 GeV mass range, and first limits on the Higgs boson production cross
section times the $H \ra \g Z^0$ branching fraction have been derived
\cite{Chatrchyan:2013vaa,ATLAS-CONF-2013-009}. The collaborations set an upper
limit on the ratio $\Gamma_{\g Z} /\Gamma_{\g Z}^{\rm SM} < 10$. A measurement
of $\Gamma_{\g Z^0}$ can also provide insights about the underlying dynamics of
the
Higgs sector since new heavy charged particles can alter the SM prediction, just
as for the $H\ra \g\g$ channel, without affecting the gluon-gluon fusion Higgs
production cross section ~\cite{Carena:2011aa}. Moreover, the
measurement of $H \ra \g Z^0$ and its rate compared to $H\ra \g\g$ is crucial
for
broadening our understanding of the EWSB pattern
\cite{Djouadi:1996yq,Djouadi:2005gi,Kniehl:1993ay}. Testing
the SM nature of this Higgs state and inspecting possible deviations in its
coupling to SM particles will represent a major undertaking of modern particle
physics and a probable probe of models going beyond the Standard Model (BSM).
\par\noi 
Despite the fact that no significant deviation from the SM has been observed,
there are many theoretical arguments and observations from
astroparticle physics and cosmology supporting the fact that it cannot be the
final answer for a complete description of Nature. If New Physics must enter the
game, what the experimental results told us so far is that any BSM should
exhibit decoupling properties. Among BSM, supersymmetry
(SUSY) is probably the best motivated and most studied framework. Its minimal
incarnation, the minimal supersymmetric standard model (MSSM), although
possessing such a decoupling regime, relies heavily on the features of the
stop sector to reproduce a 125 GeV Higgs boson; see 
\cite{Draper:2011aa,Hahn:2014,Buchmueller:2014,Buchmueller:2014a}. The
introduction of an additional gauge-singlet
superfield $S$ to the MSSM content, whose simplest version is dubbed as the
Next-to-Minimal Supersymmetric Standard Model (NMSSM)
\cite{Ellwanger:2009dp,Maniatis:2009re}, relaxes such an upper bound and alters
the dependence with respect to the stop sector. The singlet term provides an
extra tree level contribution to the Higgs
mass matrix such that the MSSM limit can be exceeded, already at the tree level
\cite{Barbieri:2006bg,Ellwanger:2006rm,Ellwanger:2011mu}. The neutral CP-even
Higgs sector is then enlarged with three states $h_i^0$, where $i$ ranges from 1
to 3 and is ordered in increasing mass. In this context the
lightest CP-even Higgs state might well be dominantly singlet with reduced
couplings to the SM and thus could remain essentially invisible at colliders:
the SM-like Higgs state would then be the second lightest and a small mixing
effect with the singlet would in turn shift its mass towards slightly higher
values. The NMSSM possesses another virtue, in addition to the ones of the MSSM:
the so-called $\mu$ problem of the MSSM \cite{Kim:1983dt} can be circumvented
as it is dynamically generated once the singlet field gets a vacuum expectation
value (vev) \cite{Ellwanger:2009dp,Maniatis:2009re}. All in all, the NMSSM
now appears as more appealing than the MSSM and has received considerable
attention
\cite{Hall:2011aa,Ellwanger:2011aa,Arvanitaki:2011ck,King:2012is,Kang:2012sy,
Cao:2012fz,Ellwanger:2012ke,Jeong:2012ma,Randall:2012dm,Benbrik:2012rm,
Kyae:2012rv,Agashe:2012zq,Belanger:2012tt,Heng:2012at,Choi:2012he,King:2012tr,
Gherghetta:2012gb,Cheng:2013fma,Barbieri:2013hxa,Badziak:2013bda,Cheng:2013hna,
Hardy:2013ywa}.
\par\noi
In the present work, we investigate the $\g Z^0$ decay channel of the SM-like
Higgs boson $h$ of the NMSSM and its correlation with $H \ra \g\g$. We compute
these two decay widths with the help of the automatic code \texttt{SloopS}
\cite{Baro:2008bg,Baro:2009gn}, initially designed to tackle one-loop
calculations in the MSSM
\cite{Boudjema:2005hb,Baro:2007em,Baro:2009ip,Baro:2009na}. This code has been
recently further developed to deal with the NMSSM extended field content and
applied to Dark Matter \cite{Chalons:2011ia,Chalons:2012hf,Chalons:2012xf} 
and Higgs phenomenology \cite{Chalons:2012qe}. Thanks to this implementation,
all the relevant particles running in the loops are properly taken into
account, in particular the non-diagonal contributions due to the
non-diagonal couplings of the $Z$ boson to charginos and sfermions. 
Our results are consistent with those presented in \cite{Cao:2013ur}.
As compared to \cite{Cao:2013ur},  we perform a more thorough 
exploration on the parameter space of the NMSSM, in particular also considering
the small $\l$ region, and we compute the expected signal strengths for both the
vector boson fusion production mode (VBF) and
the gluon fusion mode ($gg$). In addition we impose the most recent collider constraints on the Higgs sector using
\texttt{HiggsBounds}\cite{Bechtle:2013wla} and  \texttt{HiggsSignals}\cite{Bechtle:2013xfa} to require that one of the
Higgses fits the properties of the particle observed at the LHC, thus
illustrating that the most severe deviations from the SM expectations for   
 $H\to \g Z^0$ are already constrained.
We further explicitly distinguish the case where the 125 GeV Higgs is the lightest or second lightest CP-even Higgs in the NMSSM.
\par\noi
This work is organized as follows. In the first part we quickly review the
CP-even Higgs sector of NMSSM relevant for our work, and in the second part we
discuss the calculation of the $H \ra \g\g(Z^0)$ partial widths and review the
effects of SM and SUSY particles inside the loops. In the next section we
present the implementation and the numerical evaluation of the partial width
within \texttt{SloopS} and then we carry out a numerical investigation to
explore to what extent the signal can be enhanced in the NMSSM after
applying various experimental constraints. In the last section we draw our
conclusions.

\section{CP-even Higgs sector of the NMSSM}
In the NMSSM superpotential the $\mu$-term involving the two Higgs doublet
superfields $\hat H_u$ and $\hat H_d$ is  absent and a gauge singlet superfield
$\hat S$ is added instead \cite{Ellwanger:2009dp,Maniatis:2009re}:
\begin{equation}
\label{superpot}
 W_{\mathrm{NMSSM}} = W_{\mathrm{MSSM}}^{\mu=0} + \l \hat S \hat H_u \cdot \hat
H_d
\end{equation}\noi 
The MSSM $\mu$ bilinear term has now been replaced by the trilinear coupling of
the singlet with the two doublets and any dimensionful terms are forbidden by
requiring the superpotential to be $\mathbb{Z}_3$ symmetric. Once the singlet
acquires a vev $s=\langle S \rangle$, an effective $\mu$ term is generated with
respect to the MSSM, which is then naturally of the order of the electroweak
(EW) scale,
\begin{equation}
 \mu_{\mathrm{eff}} = \l s
\end{equation}\noi
 The soft-SUSY breaking Lagrangian is also modified according to
\begin{eqnarray}
 -\Lag_{\mathrm{soft}} &=& m^2_{H_u} |H_u|^2 +  m^2_{H_d} |H_d|^2 + m_S^2 |
S|^2\non \\
&+&(\lambda A_\lambda H_u \cdot H_d S + \frac{1}{3} \kappa A_{\kappa} S^3 + h.c)
\end{eqnarray}\noi 
where $H_u \cdot H_d $ stands for the usual $SU(2)$ product: $X \cdot Y = X^T
\varepsilon Y$ with $\varepsilon = -i \sigma_2$. Given $M_Z$ and using
conditions coming from the minimization of the Higgs
potential, one can choose six independent parameters for the Higgs sector 
\begin{equation}
 \lambda, \kappa, A_\lambda, A_\kappa, \mu_{\mathrm{eff}}, \tb
\end{equation}\noi 
where $\tb = \mathrm{tan} \beta = v_u/v_d$, the ratio of the two Higgs doublet
vevs: $\langle H_u^0 \rangle = v_u,\langle H_d^0 \rangle = v_d$ . From the
superpotential in Eq.~(\ref{superpot}), one derives the tree-level Higgs
potential containing the $D$-, $F$- and soft-SUSY breaking terms (we stick to
real parameters):
\begin{eqnarray}
\label{scalarpot}
V_0 &=& \left(m^2_{H_u} + \l^2| S|^2 \right)|H_u|^2 + \left(m^2_{H_d} + \l^2|
S|^2 \right)|H_d|^2 +| \l H_u \cdot H_d + \k S^2|^2\non \\
      &+&\frac{g^{'2}}{8}\left(|H_u|^2 - |H_d|^2\right)^2 +
\frac{g^2}{8}\left[\left(|H_u|^2 + |H_d|^2\right)^2 -4 |H_u\cdot H_d|^2\right]
\non \\
      &+& m_S^2 |S|^2 + \left[\l A_\l S H_u \cdot H_d+\frac{1}{3}\k A_\k S^3  +
h.c\right]
\end{eqnarray}\noi 
with $g'$ and $g$ being the $U(1)_Y$ and $SU(2)_L$ couplings respectively.
The parameters $m^2_{H_u}, m^2_{H_d}$ and $m_S^2$ can be traded for the vevs
$v_u,v_d,s$ through the minimization conditions of $V_0$. The neutral physical
fields are obtained by expanding the full scalar potential Eq. (\ref{scalarpot})
around the vev's as
\begin{equation}
\label{conventions}
  H_d = \begin{pmatrix}
         v_d + (h_d^0 + i a_d^0)/\sqrt{2} \\ - H_d^-
        \end{pmatrix},\hspace{2mm}
 H_u = \begin{pmatrix}
         H_u^+ \\ v_u + (h_u^0 + i a_u^0 )/\sqrt{2}
        \end{pmatrix},\hspace{2mm}
 S =  s + (h_s^0 + i a_s^0 )/\sqrt{2}
\end{equation}\noi 
The $3\times 3$ symmetric CP-even Higgs mass matrix is derived by collecting the
real parts and reads, in the basis $(h_u^0, h_d^0, h_S^0)$,
\begin{equation}
\label{CPevenmat} 
\mathcal{M}^2_S = 
\begin{pmatrix}
M_Z^2 \sinb^2 + \mu_{\rm eff}B_{\rm eff} c\tb   &\left(\l^2 v^2 -
M_Z^2/2\right)\sbtwo -\mu_{\rm eff} B_{\rm eff} &\l v (2 \mu_{\rm ef
f} \sinb-(B_{\rm eff} +  \nu) \cosb)\\
\cdot & M_Z^2 \cosb^2 + \mu_{\rm eff}B_{\rm eff}\tb  & \l v (2 \mu_{\rm ef
f} \cosb-(B_{\rm eff} +  \nu) \sinb)\\
\cdot & \cdot &\l^2 v^2 A_\l \sbtwo/2 \mu_{\rm eff}+\nu (A_\k + 4 \nu) 
\end{pmatrix}
\end{equation}\noi 
where we have traded the $SU(2)_L\times U(1)_Y$ gauge couplings for the gauge
boson masses through $M_Z^2 = (g^2+g^{'2})v^2/2, M_W^2 = g^2 v^2/2$. We used
the short-hand notations $\nu = \k s, B_{\rm eff}= A_\l + \nu$ as well as
$\cos \b = \cosb, \sin \b = \sinb$ and so forth for trigonometric functions. The
physical eigenstates $h_i^0$ (with $i=1$ to $3$) are obtained by nationalizing
Eq. (\ref{CPevenmat}) with an orthogonal matrix $S_h$ such that ${\rm
diag}(m_{h_1^0}^2,m_{h_2^0}^2,m_{h_3^0}^2) = S_h \mathcal{M}^2_S
S_h^{-1}$. Although it is possible, as said in the Introduction, to reproduce a
125 GeV SM-like Higgs mass in the NMSSM already at the tree-level, radiative
corrections (mostly from the stop sector) to the Higgs sector are significant
~\cite{Degrassi:2009yq,Ellwanger:2009dp,Baglio:2013iia} and should be taken into
account. For this purpose, we computed the radiatively corrected Higgs masses
using the publicly available code \texttt{NMSSMTools}
\cite{Ellwanger:2004xm,Ellwanger:2005dv,Ellwanger:2006rn}.

\section{The \texorpdfstring{$H \ra \g\g$}{H to gam gam} and \texorpdfstring{$H
\ra \g Z^0$}{H to gam Z} decay widths}
The partial width $\Gamma_{\g \g}$ and $\Gamma_{\g Z}$ in the case of the
CP-even $h_i^0$ Higgs bosons can be written in SUSY generically as, 
\begin{align}
 \Gamma_{\g\g}(h_i^0) &= \frac{\alpha^2 G_F^2 m_{h_i^0}^3}{128\sqrt{2}
\pi^3}\left|\sum_f \mathcal{A}_f^{\g\g}+ \mathcal{A}_W^{\g\g} +
\mathcal{A}_{H^\pm}^{\g\g}+\sum_{\tilde f} \mathcal{A}_{\tilde
f}^{\g\g}+ \mathcal{A}_{\chi^\pm}^{\g\g}\right|^2 \\
 \Gamma_{\g Z}(h_i^0) &= \frac{\alpha G_F^2 m_W^2 m_{h_i^0}^3}{64
\pi^4}\left(1-\frac{m_Z^2}{m_{h_i^0}^2} \right)\left|\sum_f
\mathcal{A}_f^{\g Z}+ \mathcal{A}_W^{\g Z} +
\mathcal{A}_{H^\pm}^{\g Z}+\sum_{\tilde f} \mathcal{A}_{\tilde
f}^{\g Z}+ \mathcal{A}_{\chi^\pm}^{\g Z}\right|^2
\end{align}\noi
The analytic expression for each amplitude ${\cal A}_j$ ($j=W,f,H^\pm, \tilde
f,\chi^\pm$) can be found in \cite{Djouadi:1996yq,Djouadi:2005gj,Cao:2013ur}.
For the sake of completeness, let us first discuss the SM-like contributions
$\mathcal{A}_W$ and $\mathcal{A}_f$. As the Higgs boson couples to SM particles
proportionally to their mass, its couplings to neutral gauge bosons are
dominantly mediated by the heaviest charged particles of the SM: the $W^\pm$
boson and third generation quarks ($f=t,b$). The growing of the couplings with
the mass counterbalances the decrease of the triangle amplitudes with
an increasing loop mass, thereby not decoupling the contribution of heavy
particles. The remaining fermion contributions are much smaller, due to smaller
masses. These two partial widths are therefore interesting probes of the number
of heavy charged particles which can couple to the Higgs boson. In both decay
channels the $W$ loops are by far the dominant ones and about 4.5 times larger
than the top quark amplitude for a 125 GeV Higgs bosons for the $\g\g$ channel
and about one order of magnitude in the $\g Z^0$ case \cite{Djouadi:1996yq}.
However, the total width is significantly reduced by the destructive
interference between the two contributions. The full two-loop corrections
(EW+QCD) for the SM-like Higgs decay into $\g \g$ and is under 2\%
\cite{Passarino:2007fp} below the $W^+W^-$ threshold. The complete QCD
corrections at the three loop level are also known and were presented in
\cite{Maierhofer:2012vv}. For the $\g Z^0$ decay width, the two loop QCD
corrections to top quark loops were computed in \cite{Spira:1991tj} and the
relative magnitude of the QCD correction to the partial width for a 125 GeV
SM-like Higgs boson is below 0.3 \%.
\par\noindent
In SUSY theories, the additional superpartners of the SM particles do not
couple to the Higgs boson proportionally to their masses, as the masses are
 generated through the soft-SUSY breaking Lagrangian and not the Higgs
mechanism. Hence, these contributions are suppressed by the heavy masses running
in the loops. However, if some of the superpartners' masses are not too large,
most notably the lightest chargino and third generation squarks when the
squark mixing angle is large, the decay channels can be affected and their
contribution can enable a discrimination between the lightest SUSY and standard
Higgs boson even in the decoupling regime \cite{Djouadi:2005gj}. Although the
$h_i^0\ra \g Z^0 $ partial width is generically suppressed with respect to
$h_i^0 \ra \g\g$, this channel is of interest as the non-diagonal couplings of
the Higgs and the $Z^0$ gauge boson to sfermions $\tilde f_1 \tilde f_2$ and
$\chargomp \chargtpm$ pairs are active. They are absent in the two-photon case
due to the $U(1)_{\rm QED}$ gauge invariance. The complete analytical LO SUSY
amplitudes can be found in \cite{Djouadi:1996yq,Djouadi:2005gj,Cao:2013ur}. The
common lore is that these non-diagonal contributions are ignored since they are
in general small for most purposes \cite{Djouadi:2005gj}. In the package
\texttt{NMSSMTools} \cite{Ellwanger:2004xm,Ellwanger:2005dv,Ellwanger:2006rn}
the diagonal chargino loop contributions only are included and all
sfermions contributions are missing. It was pointed out in
\cite{Cao:2013ur} that sometimes they may play a role and that some numerical
factors were missing in \cite{Djouadi:1996yq}. In the present work we computed
these partial widths with the numerical code
\texttt{SloopS}\cite{Baro:2008bg,Baro:2009gn} which
automatically generates the one-loop amplitudes and shuns hand calculation
errors thanks to internal checks like ultraviolet (UV) finiteness and gauge
invariance. We now turn to the description of the numerical computation in the
next section. 

\section{Numerical evaluation of \texorpdfstring{$\Gamma_{\g\g}$}{H gam gam}
and \texorpdfstring{$\Gamma_{\g\g}$}{H gam Z} with \texttt{SloopS}}
In \texttt{SloopS}, the complete spectrum and set of vertices are generated at
the tree level through the LanHEP package
\cite{Semenov:1997qm,Semenov:1998eb,Semenov:2010qt}. The complete set of
Feynman rules is then derived automatically and passed to the bundle
\texttt{FeynArts}/\texttt{FormCalc}/\texttt{LoopTools}
\cite{Hahn:1998yk,Hahn:2000kx,Hahn:2004rf}(that we denote as \texttt{FFL}). A
powerful feature of \texttt{SloopS} is the ability to check not
only the UV finiteness check as provided by \texttt{FFL}, but also the gauge
independence of the result through a generalized gauge fixing Lagrangian, which
was adapted to the NMSSM \cite{Chalons:2011ia}. The gauge-fixing Lagrangian can
be written in a general form:
\begin{equation}
\label{gaugefixing}
{\mathcal L}_{GF} = -\frac{1}{\xi_W} F^+ F^- - \frac{1}{2  {{ \xi_Z}}  }|
F^Z|^2 - \frac{1}{2 {{ \xi_A}} } | F^A|^2
\end{equation}\noi 
For the $\Gamma_{\g\g}$ and $\Gamma_{\g Z}$ decay width only the
nonlinear form of $F^\pm$ is of relevance and is given by\footnote{The other
gauge-fixing functions $F^Z$ and $F^A$ can be kept in the usual $R_\xi$ form.
The complete expressions can be found in \cite{Chalons:2011ia}. For
practical purposes we set $\xi_{W,Z,A} =1$.}
\begin{equation}
F^+  =  \bigg(\partial_\mu - ie {\tilde{\alpha}}  A_\mu - igc_W 
{\tilde{\beta}} Z_\mu\bigg) W^{\mu \, +}
+ i{{ \xi_W}}  \frac{g}{2}\bigg(v +  \sum_{i=1}^3{\tilde{\delta}}_i
h_i^0  \bigg)G^+
\end{equation}\noi
The parameters $\tilde \alpha,\tilde \beta$ and $\tilde \delta_i$ are dubbed as
nonlinear gauge (NLG) parameters and $G^\pm$ are the charged Goldstone fields.
We recover the usual 't Hooft-Feynman gauge by setting these parameters to
vanishing values. The ghost Lagrangian ${\cal L}_{Gh}$ is established by
requiring that the full Lagrangian is invariant under BRST
transformations.\footnote{The BRST transformations for the gauge fields can be
found in \cite{Belanger:2003sd} and for the scalar fields in
\cite{Chalons:2011ia}.}. The gauge dependence is in
turn transferred from the vector boson propagators to a modification of the
ghost-Goldstone-vector boson vertices (see for example \cite{Belanger:2003sd}).
Numerically the gauge invariance check is performed by varying the
parameters $\tilde\alpha,\tilde\beta$ and $\tilde\delta_i$. 
\par\noi 
Similarly to the MSSM, radiative corrections to the Higgs masses in the NMSSM
can be relatively large (see e.g \cite{Degrassi:2009yq} and references therein)
and thus significantly affect the kinematics of the decay. As
said previously, we used \texttt{NMSSMTools} to compute the Higgs spectrum and
rotation matrices. Since the Higgs spectrum and parameters entering the Higgs
potential in Eq. (\ref{scalarpot}) are not independent quantities, a shift on
the tree level Higgs masses also results in a shift of the parameters and
thus on the Higgs-to-Higgs couplings. Therefore, to parametrize the radiative
corrections to the Higgs masses and couplings we make use of the effective
Lagrangian devised in \cite{Chalons:2012qe} and given by
\begin{equation}
\label{effpot}
 V_{\rm eff} = V_0 + V_{\rm rad}
\end{equation}
where $V_0$ is the same as Eq.(\ref{scalarpot}) and
\begin{eqnarray}
 \label{Z3potLH}
V_{\rm rad}
& = &
\frac{{\lambda}_1}{2}
|H_d|^4+\frac{\lambda_2}{2}
|H_u|^4+\lambda_3|H_u|^2|H_d|^2+\lambda_4|H_u\cdot H_d|^2
+{\bar\kappa}^2|S|^4+\frac{1}{3}\left(\bar{A}
_SS^3+h.c.\right) \non \\
& &+{\lambda}_P^u|S|^2|H_u|^2+{\lambda}_P^d|S|^2|H_d|^2
+\left[{A}_{ud } SH_u\cdot H_d+{\lambda}_P^MS^{*\,2}H_u\cdot
H_d+h.c.\right]
\end{eqnarray}\noi 
Once the spectrum and mixing matrices are known from \texttt{NMSSMTools} we
solve for the $\l$s\footnote{We denote generically as ``$\l$'' any parameter
entering Eq. (\ref{Z3potLH}).} using equations derived from the effective Higgs
mass matrices extracted from Eq. (\ref{effpot}). We refer to
\cite{Chalons:2012qe} for further details concerning the extracting procedure of
the $\l$ parameters. This procedure then ascertains the gauge independence
of the computation. This was explicitly demonstrated analytically and
numerically in \cite{Chalons:2012qe} for the partial width $h_i^0 \ra \g\g$.
The nonlinear gauge-fixing Lagrangian possesses another virtue: setting
particular values to the NLG parameters can cancel specific vertices. For the
case at hand, with the peculiar choice $\tilde \alpha = -1$ the coupling $\g
W^\pm G^\mp$ is absent due to an underlying $U(1)_{\rm QED}$ symmetry
conserving gauge-fixing function $F^\pm$ and less diagrams have to be
considered. This property is therefore a
welcomed feature and was employed to simplify the analytic calculation of $H \ra
\g \g$ in \cite{Shifman:1979eb,Gavela:1981ri}. Although in such a gauge the
calculation of the SM-like amplitude of $\Gamma_{\g Z}$ is not easily
translated from $\Gamma_{\g \g}$, since the vanishing of the $\g W^\pm G^\mp$
vertex introduces an asymmetric treatment of the photon and the $Z^0$ boson
 \cite{Bergstrom:1985hp}, the evaluation of the $Z -\g$ transition diagram
$\Pi_{\g Z}(0)$ is not needed. Indeed, for $\tilde \alpha =-1$ this mixing
self-energy vanishes at $q^2=0$ thanks to, once more, the fact that $F^\pm$
preserves the $U(1)_{\rm QED}$ gauge symmetry \cite{Belanger:2003sd}. This is of
importance for our numerical evaluation of $\Gamma_{\g Z}$ since we do not
generate diagrams with external wave function correction and the introduction of
the field-renormalization constant $\delta Z^{1/2}_{Z \g}=-\Pi^T_{\g
Z^0}(0)/M_Z^2$ would not be needed in such a gauge. In a general nonlinear gauge
this transition is crucial to maintaining the UV finiteness and the gauge
invariance of the result. We thoroughly checked this feature
numerically by varying the NLG parameters $\tilde \alpha,\tilde \beta$ and
$\tilde \delta_i$.

\section{Numerical investigation}

Although the properties of the Higgs boson observed at the LHC are compatible with the SM predictions~\cite{Aad:2012tfa,Chatrchyan:2013lba},
the precise measurements of all its decay channels can give some crucial information on new physics. 
In this analysis we examine the expectations for the decay $h_i^0\rightarrow
\g Z^0$ in the framework of the NMSSM after taking into account the
constraints on the Higgs observed at the LHC. In particular we quantify the
importance of the sfermions and non-diagonal charginos contributions discussed
in the previous section. 
For this we explore the parameter space of the NMSSM with emphasis on the regions which could lead potentially to large corrections, 
those with light charginos and/or light stop.
\par\noi
The chargino mass matrix is given by,
\begin{eqnarray}
\begin{pmatrix}
M_2&g v_u\\
g v_d&\mu_{eff}
\end{pmatrix}
\end{eqnarray}
while  the stop mass matrix in the $(\tilde t_R, \tilde t_L)$ basis, reads
\begin{eqnarray}
\begin{pmatrix}
m_{U_3}^2+h_t^2v_u^2-(v_u^2-v_d^2)\frac{g^{'2}}{3}&h_t(A_tv_u-\mu_{eff}v_d)\\
h_t(A_tv_u-\mu_{eff}v_d)&m_{Q_3}^2+h_t^2v_u^2+(v_u^2-v_d^2)\left(\frac{g^{'2}}
{12}-\frac{g^2}{4}\right)
\end{pmatrix}
\end{eqnarray}
where $M_2$ is the SU(2) gaugino mass, 
$ m_{Q_3}, m_{U_3}$ are the soft masses for the stops and $A_t$ is the stop
trilinear mixing. Since $h_t$ is of the order 1, the mixing between both stops
can be important.  The large mixing can lead to  large radiative corrections  to
the SM-like Higgs mass and to one of the  stops being quite light. Thus, the
main squark contribution to the Higgs loop-induced decays ($\gamma\gamma, \gamma
Z^0$ and $gg$) is coming from the stop sector. 
\par\noi
To restrict the number of free parameters of the phenomenological NMSSM, 
we perform a  scan over only the  parameters most relevant for the Higgs mass
(Eq.
\ref{CPevenmat}) and couplings, specifically those   
of the chargino, squark and Higgs sectors  which we take in the following range:
\begin{eqnarray}
100~\textrm{GeV}<&\mu &<500 \textrm~{GeV}\notag\\
100~\textrm{GeV}<&M_2&<1000 \textrm~{GeV}\notag\\
0~\textrm{GeV}<&t_{\beta}&<20\notag\\
0<&\lambda,\kappa &<0.7\notag\\
100~\textrm{GeV}<&A_{\lambda}&<1000~\textrm{GeV}\notag\\
-1000~\textrm{GeV}<&A_{\kappa}&<-100~\textrm{GeV}\notag\\
-3000~\textrm{GeV}<&A_t&<3000~\textrm{GeV}\notag\\
400~\textrm{GeV}<&m_{\tilde Q_3},m_{\tilde U_3}&<2000~\textrm{GeV}\notag
\end{eqnarray}
We assume that all  squarks of the first and second generations are heavy ($m_{\tilde 
Q_i}=m_{\tilde u_i}=m_{\tilde d_i}=2~{\rm TeV}$) 
as well as the right-handed sbottom mass, $m_{\tilde d_3}=2~{\rm TeV}$,
since they do not play an important role in Higgs physics. We also 
assume that all sleptons are heavy, $m_{\tilde L_i}=m_{\tilde l_i}=2~{\rm TeV}$
\footnote{We have made additional scans to check the impact of varying the
parameters of the stau sector. Corrections to $h_i^0\rightarrow \g Z^0$ lie
below a few percent except for very light staus (below the LEP limit). We expect
that for values of $\tan\beta$ much larger than those considered here, one can
get large enhancement to the $\gamma Z^0$ branching ratio as was shown
previously for the $h\rightarrow \gamma\gamma$ in the
MSSM~\cite{Carena:2011aa}.},
as well as the gluino, $M_3=1.5~{\rm TeV}$. The most important LHC constraints
on supersymmetric particles are then automatically satisfied. Finally, we set
$M_1=150~{\rm GeV}$, the exact value of the neutralino LSP is not very important
for our analysis, provided the neutralino is too heavy for the Higgs to decay
invisibly. However, the value of $M_1$ could be adjusted to ensure that the
limit on the stop mass from the LHC is satisfied (the limit on the lightest stop
can  easily be avoided when $m_{\tilde t_1}-m_{\tilde\chi_1}<
m_t$)~\cite{ATLAS-stop,Chatrchyan:2013xna}. Note that we concentrate on small
values of $\tan\beta$ since it is  the region  where the Higgs sector can differ
significantly from that of the MSSM. In this region one can find large
deviations in the $h_i^0\rightarrow \gamma\gamma$ decay ~\cite{Ellwanger:2011aa}
due in particular to the singlet component of the Higgs. Thus, large deviations
are also expected for $h_i^0\rightarrow \g Z^0$.
\par\noi
In the NMSSM, either of the light scalar, $h_1^0$ or $h_2^0$, could be the one
observed at the LHC with a mass of 125 GeV:  we consider both possibilities.  We
use \texttt{NMSSMTools} to compute the supersymmetric spectrum and to impose
constraints on the parameter space\footnote{See also \cite{Baglio:2013iia} for
a full one-loop calculation of the Higgs boson spectrum in the real and complex
NMSSM using a mixed DR and OS renormalization scheme and for a comparison
between the different methods.}, specifically:
the large electron positron (LEP) collider constraints on Higgs and chargino
masses as well as the constraint that there be no unphysical global
minimum on the Higgs potential. We select only the points with one Higgs in the mass range $122-128$ GeV. 
Finally, we impose the  collider constraints on the Higgs sector  from
\texttt{HiggsBounds} \cite{Bechtle:2013wla} and require that one Higgs fits the
properties of the particle observed at the LHC using \texttt{HiggsSignals}
\cite{Bechtle:2013xfa}.  For
these two codes, we choose a theoretical uncertainty of 2 GeV for the Higgs
masses. The allowed points are those for which $\chi^2< \chi^2_{\rm best
fit}+18.3$ corresponding to the 95\% confidence level (C.L) for ten free
parameters. Note that the best fit point is slightly better than the SM. 
\par\noi 
First, we checked  whether sfermions and non-diagonal charginos contributions 
have a significant impact on the the process $h_i^0\ra \g Z^0$;  for this we
calculate the
following ratio :
\begin{equation}
R=\frac{\Gamma(h_i^0\rightarrow \gamma Z^0)_{total}}{\Gamma(h_i^0\rightarrow
\gamma
Z^0)_{restricted}}\quad ,
\end{equation}
where $\Gamma(h_i^0\rightarrow \gamma Z^0)_{total}$ is the decay width
calculated
with all the possible particles in the loops, and $\Gamma(h_i^0\rightarrow
\gamma
Z^0)_{restricted}$ is the one calculated by omitting sfermions and non-diagonal
charginos contributions in the loop. 
\par
\begin{figure}[h!]
   \centering
\includegraphics[width=0.49\textwidth]{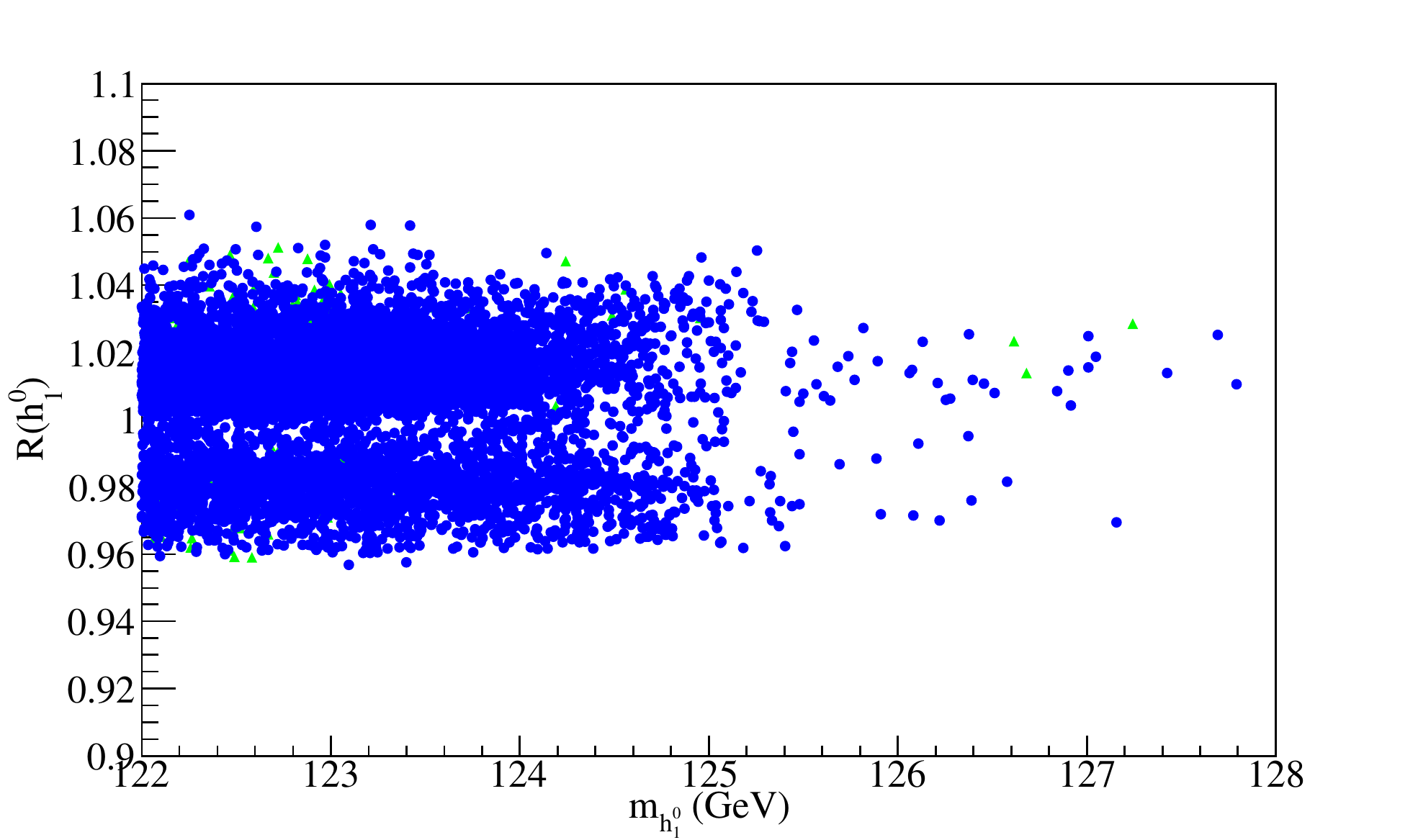} 
\includegraphics[width=0.49\textwidth]{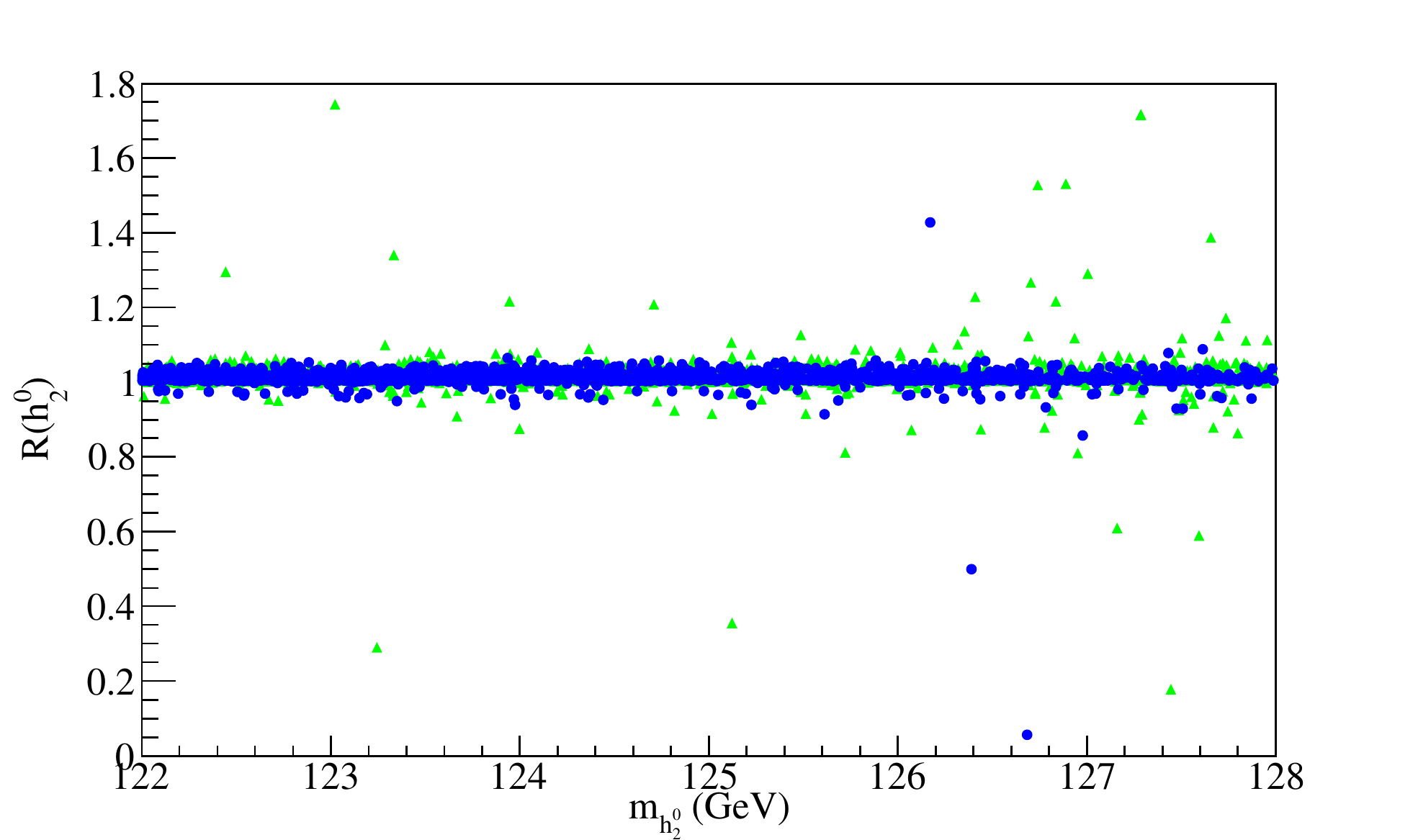}
    \caption{Ratio of the decay width of $h_1^0\ra \g Z^0$ (left panel) and
$h_2^0\ra \g Z^0$ (right panel) with all possible
particles in the loop by the one without sfermions nor non diagonal Z-charginos
contributions as a function of the mass of $h_i^0$. Blue points are allowed and
green triangles are excluded 
by LHC constraints on the Higgs.}\label{h1GamZ}
\end{figure}

\noi
The results for the ratio are shown in Fig. \ref{h1GamZ}. In
both cases, the ratio is plotted as a function of the mass of the corresponding
Higgs boson. We notice that, for most of the points, the effect is less
than $\sim 10\%$. In fact, the main effect is coming from the chargino
contributions;  we have checked that with only the sfermionic contribution the
effect is less than 5\%. 
However, for a few points, the variation can be as high as
$70\%$ in the case of $h_2^0$. These large deviations are found for points for which $h_2^0$
is almost singlet : in this case the partial decay width is suppressed since the W contribution becomes negligible and the chargino (higgsino) contribution can become relatively more important. Note that for the singlet case
the total decay width is also suppressed, leading to a branching ratio that can
be either enhanced or suppressed relative to the SM. 
However, most of these points are excluded by LHC constraints on the Higgs
sector.

\par\noi
We next consider the expectations for the $h_i^0\rightarrow \gamma
Z^0,\gamma\gamma$ branching ratios in the NMSSM as compared to the SM when
including all contributions. 
The case where $h_1^0$ is near 125 GeV leads to only mild variations of the
$\gamma Z^0$ branching ratio - typically $\approx 10\%$. In a few cases however
deviations as large as 25\% can be found; typically they are found when $h_1^0$
has a significant singlet component.
In all cases we found  a strong correlation to the $h_1^0 \rightarrow
\gamma\gamma$ branching ratio (within 10\%).  The two-photon mode should
therefore provide a better probe of new physics effects in the NMSSM considering
it can be measured with a much better precision.  
\par
\begin{figure}[h!]
   \centering
    \includegraphics[width=0.49\textwidth]{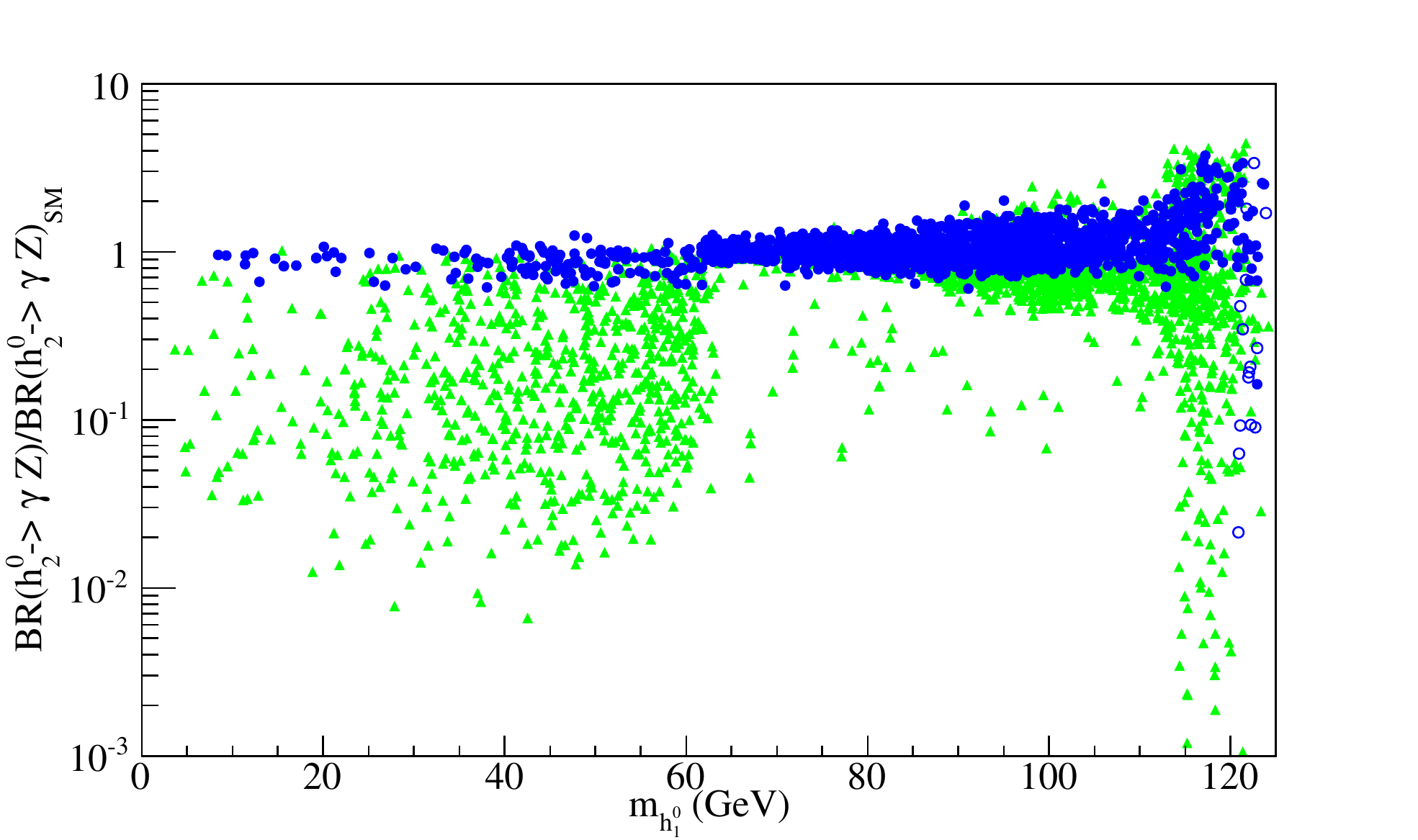}
    \includegraphics[width=0.49\textwidth]{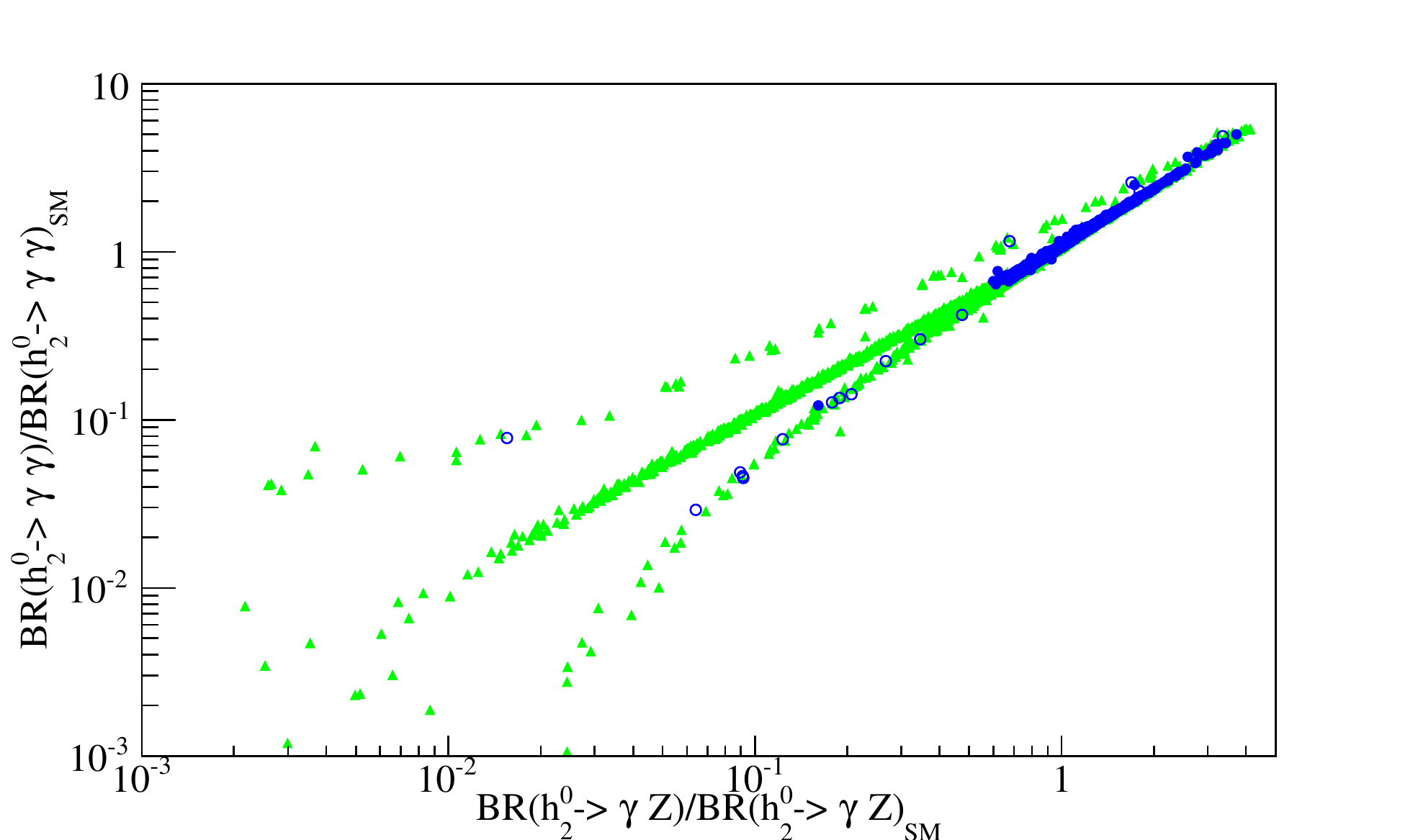}
    \caption{Left panel : Branching ratio of $h_2^0$ in $\gamma Z^0$ in the
NMSSM
relative to the SM as a function of the mass of $h_1^0$. The green triangles
correspond to points excluded at the 95\% C.L. by HiggsBounds and
\texttt{HiggsSignals}. The empty blue points correspond to $Sh_{23}>0.9$. Right
panel : Correlation between the reduced branching ratios $h_2^0\rightarrow\gamma
Z^0$ and $h_2^0\rightarrow\gamma\gamma$. 
}\label{RatioBRh2vsmh1}
\end{figure}

\noi
The results for $h_2^0\rightarrow \gamma Z^0,\gamma\gamma$ are more
interesting. 
Figure \ref{RatioBRh2vsmh1} shows the branching ratio $h_2^0\rightarrow \gamma
Z^0$ as compared to the SM value as  a function of the mass of the lightest
Higgs boson (recall that for these points $122 {\rm GeV}< m_{h_2^0}<128 {\rm
GeV}$). Although the bulk of the points are centered around the SM value, large
deviations can occur, in particular when the lightest Higgs is near 120 GeV. In
this case $h_2^0$  has a large  singlet component and the $Br(h_2^0\rightarrow
\gamma Z^0)$ can be up to 4 times larger than the SM or suppressed by more than
two orders of magnitude. Note that most of these points are excluded by the LHC
constraints as implemented in \texttt{HiggsSignals} either  because
the production of the singlet Higgs deviates significantly from the SM and/or 
the corresponding $\gamma\gamma$ channel which is correlated with $\gamma Z^0$ 
is incompatible with current measurements. Nevertheless, we found  few points
that satisfy the \texttt{HiggsSignals} constraints even though the $h_2^0$ is
almost a pure singlet  and thus has non SM couplings; see the empty circles in
Fig.~\ref{RatioBRh2vsmh1} which have $S_{h_{23}}>0.9$. The reason why such
points avoid the LHC constraints is that they correspond to cases where 
$h_1^0$ and $h_2^0$  are almost degenerate and the  superposition of the signal
of both Higgs bosons is what is observed at the LHC ($h_1^0$ is mainly doublet
and SM-like). 
\par\noi
The branching ratio $h_2^0\rightarrow \gamma Z^0$ can also be strongly
suppressed
when $m_{h_1^0}<60~{\rm GeV}$ and a new decay channel, $h_2^0\rightarrow h_1^0
h_1^0$, opens up, thus significantly increasing the total width;
see Fig.~\ref{RatioBRh2vsmh1}. Points with a large suppression are however
incompatible with the LEP and LHC constraints. 
\par\noi
Finally, we comment on the correlation between the $h_2^0\rightarrow \gamma Z^0$
and $h_2^0\rightarrow \gamma \gamma$ channels displayed in
Fig.~\ref{RatioBRh2vsmh1}, right panel.
As for the case of $h_1^0$, both channels are strongly correlated. The
correlation breaks down when $h_2^0$ has a strong singlet component 
and mainly for points with suppressed $\gamma Z^0/\gamma\gamma$ branching ratios
that are to a large extent excluded by LHC constraints.  

\par
\begin{figure}[h!]
   \centering
    \includegraphics[width=0.49\textwidth]{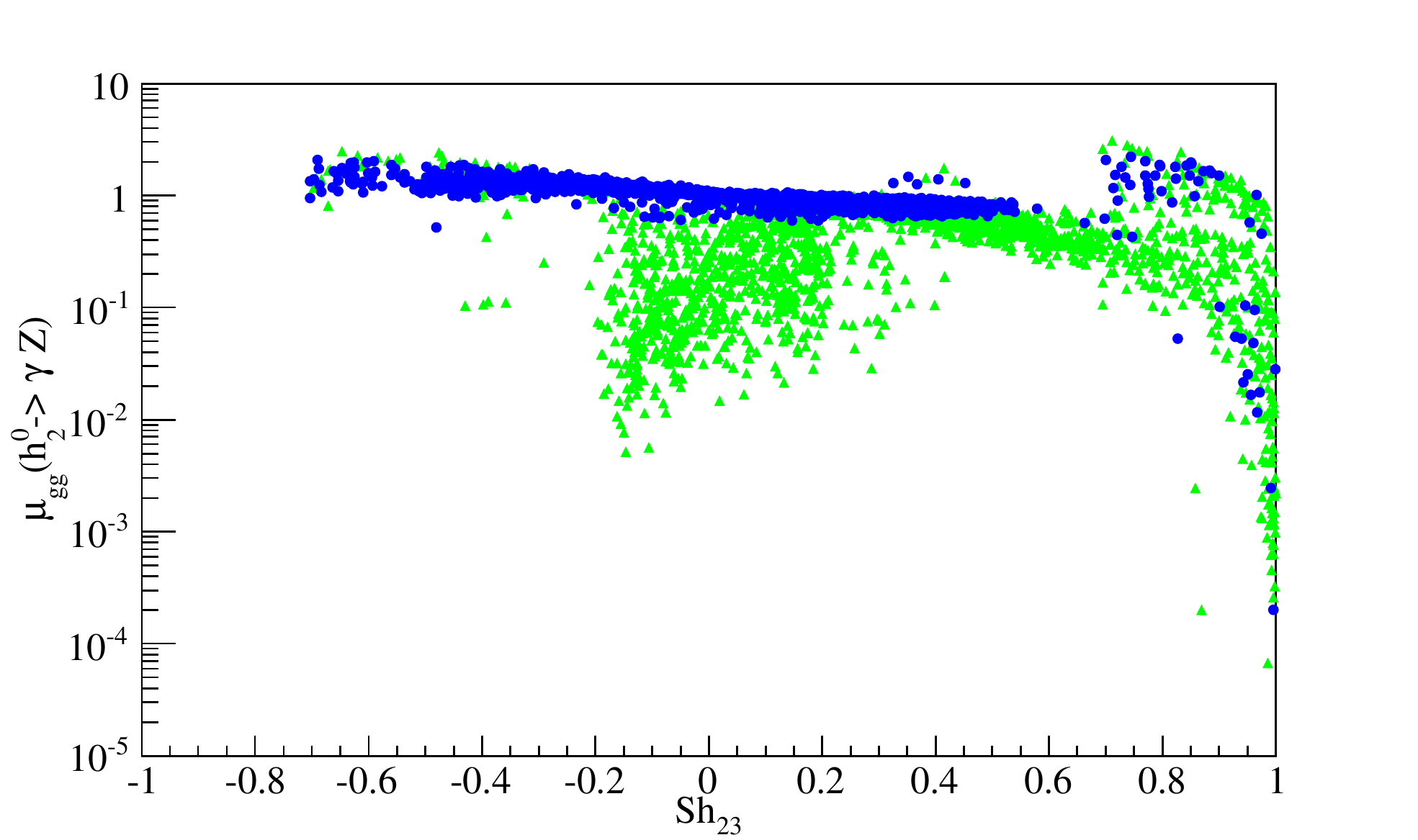}
     \includegraphics[width=0.49\textwidth]{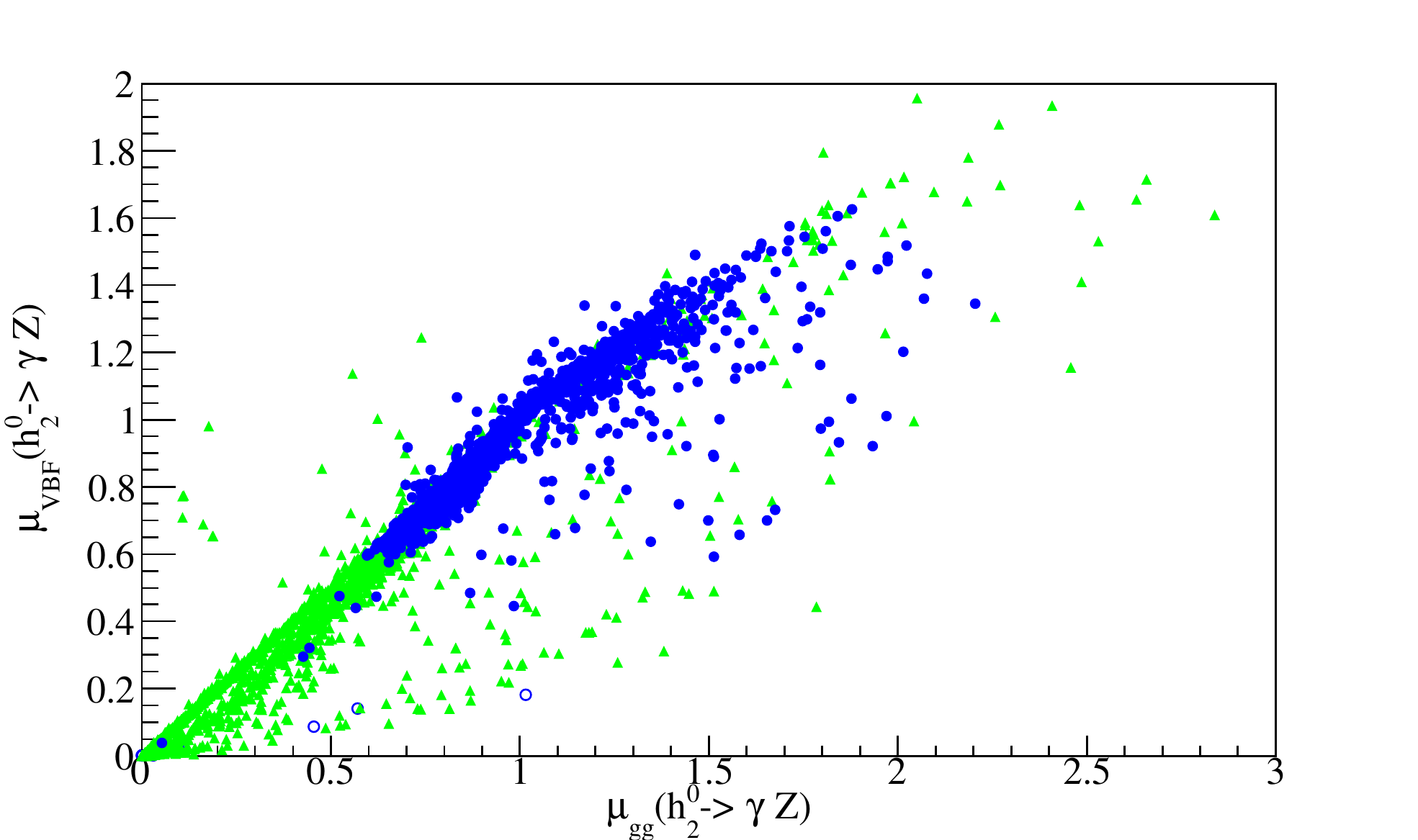}
    \caption{Left panel : Signal strength for $h_2^0\rightarrow\gamma Z^0$ for
gluon gluon fusion production with respect to $Sh_{23}$. Right panel:
Correlation between the gg and VBF signal strengths. 
Same color code as Fig.~\ref{RatioBRh2vsmh1}}\label{mugluglu}
\end{figure}

\noi
The LHC collaborations will not directly measure the branching ratio
but the signal strengths ($\mu$)  for gluon fusion (gg) and vector boson fusion
(VBF) production modes in the $\gamma Z^0$ channel. The predictions for the
signal strength in the gluon fusion mode are shown in Fig. \ref{mugluglu} for
$h_2^0$ as a function of the scalar mixing $Sh_{23}$. Three regions present an
enhancement as compared to the SM; they correspond to  (a) $S_{h_{23}}<-0.5$,
(b) $S_{h_{23}} > 0.7$ and (c) $S_{h_{23}} \approx 0.4$.
In all three cases, $h_2^0$ has  significant singlet and doublet components and
its couplings to u-type quarks and gauge bosons are somewhat suppressed while
those couplings to d-type quarks and leptons are strongly suppressed. As a
result, the total width of $h_2^0$ is much reduced and the branching ratio into
$WW,ZZ,\gamma\gamma,gg$ and $\gamma Z^0$ are all larger than in the SM. 
The signal strength in the VBF mode is mostly correlated with that in the gluon fusion mode, although 
for some points the VBF is suppressed by more than a factor 2 as compared to the
gluon fusion mode; see Fig. \ref{mugluglu} right panel. In particular, note 
that the enhancement in the gluon fusion mode ($\mu_{gg}>1$) is more important
than in the VBF mode. We had also mentioned that  for $Sh_{23}>0.9$ which
corresponds to a $h_2^0$ that is dominantly singlet, the $h_2^0\rightarrow
\gamma Z^0$ branching ratio could still be enhanced; however, the singlet
coupling to gauge bosons becomes very small  so that the VBF production mode is
suppressed and to a lesser extent also the gluon fusion production mode, making
it difficult to probe this dominantly singlet Higgs at the LHC.  

\section{Conclusion}

We have performed a complete computation of the branching ratio into $\gamma
Z^0$ of both of the two lightest CP-even Higgs in the NMSSM using
\texttt{SloopS}. Compared to previous work \cite{Cao:2013ur}, in our scans we
did not consider the large lambda regime only, but the full range of lambda
values not producing a Landau pole before the grand unified theory scale.
Moreover, we investigated to what extent the signal can be enhanced knowing the
measurements of the LHC on the Higgs boson signal strength. Also, strictly
concerning the calculation of the decay rate, we made use of a modified Higgs
boson potential to be able to use the corrected Higgs boson masses in the
kinematics of the process while maintaining gauge invariance; this was checked
thoroughly thanks to the implementation of a non-linear gauge fixing
\cite{Chalons:2012qe}. We found that the previously neglected contributions of
charginos and stops (as well as staus) generally lie below 10\%. Exploring the
parameter space of the NMSSM, we found that $h_1^0\rightarrow \gamma Z^0$ did
not vary much from the SM expectations while regions with a large enhancement
(suppression) of $h_2^0\rightarrow \gamma Z^0$ were possible, especially when
$h_2^0$ had a significant singlet component. However, most of these scenarios
are constrained by measurements of the 125 GeV Higgs boson at the LHC since the
singlet component significantly changes the coupling of $h_2^0$ to gauge bosons
and fermions (in particular b quarks).
We conclude that given the correlation between the $\gamma\gamma$ and $\gamma
Z^0$ branching ratios expected in the NMSSM,
 a better measurement of the former at the next LHC run - together with a higher
precision on other standard decay channels - will further constrain the range of
values expected for $h\rightarrow \gamma Z^0$.
Nevertheless an independent measurement of $h\rightarrow \gamma Z^0$ is useful
in probing BSM physics; for example, a  large deviation from the SM
expectations not correlated with a similar deviation in the $\gamma\gamma$ mode
would put strong constraints on the NMSSM (and other MSSM-like models).
Furthermore, a suppressed signal strength in the VBF mode relative to the gluon
fusion mode is characteristic of the partially singlet Higgs in the NMSSM.

\section{Acknowledgements}
We thank Fawzi Boudjema, B\'eranger Dumont, Guillaume Drieu La Rochelle, Alexander Pukhov  and Chris Wymant for useful discussions. 
We also thank Tim Stefaniak for providing useful information on
\texttt{HiggsSignals}. Partial funding from  the French ANR, project DMAstroLHC,
ANR-12-BS05-0006, and by tge European Commission through the HiggsTools Initial
Training Network, Grant No. PITN-GA-2012-316704, is gratefully acknowledged. The
work of G.C is supported by the
Theory-LHC-France initiative of the CNRS/IN2P3.
\newpage

\end{document}